\documentclass[12pt]{article}
\usepackage{epsfig}

\def\beq{\begin{equation}}
\def\eeq#1{\label{#1}\end{equation}}
\def\eeqn{\end{equation}}

\def\beqa{\begin{eqnarray}}
\def\eeqa#1{\label{#1}\end{eqnarray}}
\def\eeqan{\end{eqnarray}}

\let\bar=\overbar

\def\Dslash{\not{\hbox{\kern-4pt $D$}}}
\def\dslash{\not{\hbox{\kern-2pt $\del$}}}

\def\msb{{\bar{\ssstyle M \kern -1pt S}}}

\def\Title#1{\begin{center} {\Large {\bf #1} } \end{center}}

\def\modeI {B^- \to J/\psi K^-}
\def\modeII {B^0 \to J/\psi K^0}
\def\modeIII {B^- \to J/\psi K^-_1(1270)}
\def\modeIV {B^0 \to J/\psi K^0_1(1270)}
\def\modeV {B^- \to \psi  (2S) K^-}
\def\modeVI {B^0 \to \psi (2S) K^0}
\def\modeVII {B^- \to \chi_{c1} K^-}
\def\modeVIII {B^0 \to \chi_{c1} K^0}
\def\modeIX {B^0 \to \chi_{c1} K^{*0}}
\def\modeX {B^- \to J/\psi \pi^-}
\def\modeXI {B^0 \to J/\psi \pi^0}

\def\modeXII {\psi(2S) \to l^+ l^-}
\def\modeXIII{\psi(2S) \to J/\psi \pi^+\pi^-}

\begin{document}

\Title{$B$ to charmonium - mini-summary}

\bigskip\bigskip


\begin{raggedright}  

{\it Fang Fang \index{Fang, F.}\\
Department of Physics and Astronomy\\
University of Hawaii at Manoa\\
Hawaii, USA}
\bigskip\bigskip
\end{raggedright}

\section{Introduction}

I summarize recent experimental results on $B$ to
charmonium decays. Decays of $B$ meson to final states that include
charmonium states play an important role in the study of CP violation at
B-factories. The decay modes $B_{CP} \to J/\psi K_S$, $\psi (2S) K_S$,
$\chi_{c1} K_S$, $\eta_{c} K_S$, $J/\psi K_L$ and $J/\psi
K^{*0}$($K^{*0}\to K_S \pi^0$) have been used for $\rm{sin}2\phi_1$
measurements~\cite{Aubert:2001nu}~\cite{Abe:2001xe}~\cite{Aubert:2002rg}~\cite{Abe:2002wn}~\cite{Aubert:2002gv}.
These two-body decay modes are dominated by the color suppressed $b \to c$
transition. Other CP eigenstates of the neutral $B$ meson, e.g. 
$J/\psi \rho^0$, may also be useful for the CP
measurements. Meanwhile, the branching fractions for the $B \to
charmonium$ decays can provide valuable information on their decay mechanism.  

\section{Non-factorizable decay modes}

In the factorization approximation, the production of $\chi_{c0}$ and
$\chi_{c2}$ are not allowed by angular momentum and vector-current
conservation. However, these decays can occur if factorization is
broken by an exchange of soft gluons between the quarks.

\subsection{Observation of $B^+ \to \chi_{c0} K^+$}

Using a data sample containing 31.3 million $B\bar B$ events collected
at the $\Upsilon(4S)$ resonance with the Belle detector at the KEKB asymmetric
$e^+e^-$ collider, the Belle collaboration has made the first
observation of $B^+ \to \chi_{c0} K^+$~\cite{Abe:2001mw}. 

\begin{figure}[t]
\begin{center}
\epsfig{file=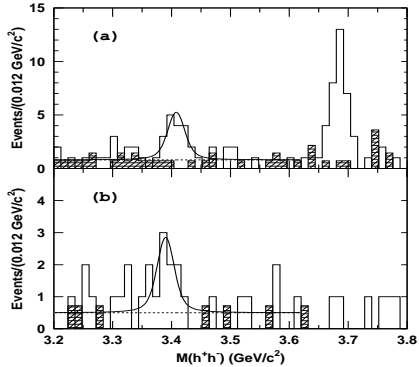,height=2in,width=2.2in}
 \caption{The (a) $\pi^+\pi^-$ and (b) $K^+K^-$ invariant mass
 spectra. Open histograms correspond to events from the $B$ signal
 region, and hatched histograms correspond to events from the $\Delta
 E$ sidebands. The curves are results of fits~\cite{Abe:2001mw}.}
\label{fig:chic0}
\end{center}
\end{figure}

The $\chi_{c0}$ candidates are reconstructed from $\chi_{c0} \to \pi^+
\pi^-$ and $K^+ K^-$. Two kinematic variables,
the beam-constrained mass, $M_{\rm bc}=\sqrt{E_{\rm
beam}^2-\vec{P}_{\rm recon}^2}$, and energy difference $\Delta E=
E_{\rm recon}-E_{\rm beam}$ in the $\Upsilon(4S)$ center of mass
frame, are formed to isolate the signal. Here $E_{\rm beam}$,
$E_{\rm recon}$ and $\vec{P}_{\rm recon}$ are the beam energy, the
reconstructed energy, and the reconstructed momentum of the signal
candidate, respectively. Figure~\ref{fig:chic0} shows the invariant
masses of
$\pi^+ \pi^-$ and $K^+ K^-$. The peaks near 3.4 $\rm{GeV}/c^2$ are
identified as the $\chi_{c0}$ meson. The peak position in the $K^+K^-$
spectrum is shifted. This could be explained by the interference of
$B^+\to \chi_{c0} K^+$ with the non-resonant $B^+\to K^+K^+K^-$. The
peak at 3.69 $\rm{GeV}/c^2$ in the $\pi^+ \pi^-$ spectrum is due to
$B^+\to \psi(2S)K^+$, $\psi(2S)\to \mu^+\mu^-$ with the muons
misidentified as pions. 

Using the $\chi_{c0} \to \pi^+ \pi^-$ decay channel,  the ratio of
branching fractions is found to be:
\[ \frac{{\cal{B}}(B^+\to \chi_{c0}K^+)}{{\cal{B}}(B^+\to J/\psi K^+)}=
 0.60^{+0.21}_{-0.18}\pm0.05\pm0.08,\]
where the first error is statistical, the second is systematic, and
the third is due to the uncertainty in the branching fraction for
$\chi_{c0}\to\pi^+\pi^-$.  
The branching fraction is measured to be
 \[{\cal{B}}(B^+\to \chi_{c0}K^+)=(6.0^{+2.1}_{-1.8}\pm1.1)\times10^{-4},\]
which is comparable to those for $B^+\to J/\psi K^+$ and
$B^+\to \chi_{c1}K^+$ decays. The $\chi_{c0} \to K^{*0} K^- \pi^+$
decay channel has been also studied and the results are in good
agreement with those determined from $\chi_{c0} \to \pi^+ \pi^-$. The
statistical significance of the signal is $6\sigma$ when these two
channels are combined. This measurement indicates a significant
non-factorizable contribution in $B$ to charmonium decays.

\subsection{Observation of $B \to \chi_{c2} X$}

The Belle collaboration has also observed $\chi_{c2}$
production in $B$-meson decay~\cite{Abe:2002wp}. The analysis is based on a
data sample containing 31.9 million $B\bar B$ events collected at the
$\Upsilon(4S)$ resonance with the Belle detector at the KEKB asymmetric
$e^+e^-$  collider. 

\begin{figure}[t]
\begin{center}
\epsfig{file=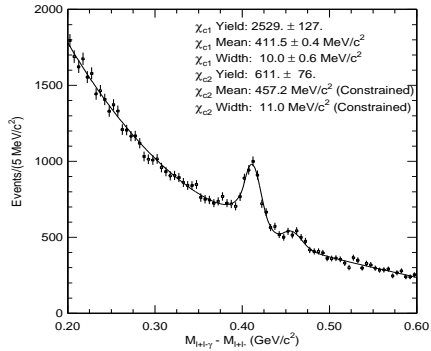,height=1.8in,width=2.2in}
 \caption{The distribution of the mass difference between the $\chi_c$
 and the $J/\psi$ candidates events~\cite{Abe:2002wp}.} 
\label{fig:chic2}
\end{center}
\end{figure}

The $\chi_{c2}$ candidates are reconstructed via $J/\psi\gamma$,
$J/\psi \to l^+l^-$. The photon energy resolution is studied using
$D^{*0} \to D^0 \gamma$ decay. The $\sigma _E$/$E_{\gamma}$ is
($2.61\pm0.04$)\% around 400 $\rm{MeV}$. The good photon energy
resolution leads to a clear separation between the $\chi_{c2}$ peak and the
larger $\chi_{c1}$ peak as shown in Fig.~\ref{fig:chic2}. To extract
signal yields, the distribution
is fit to two Crystal Ball line shapes and a third-order Chebyshev
polynomial for the background. The continuum
subtracted yield for $B \to \chi_{c2}$ is $607^{+76}_{-94}$ events and the
branching fraction for $B \to \chi_{c2} X$ is
($1.80^{+0.23}_{-0.28}\pm0.26$) $\times$ $10^{-3}$. The branching
fraction for $B \to \chi_{c1} X$ is also measured. The momentum
spectrum of $\chi_{c2}$ does not show a significant contribution from two body
$B \to \chi_{c2} K$, in contrast to the momentum spectrum for $B \to \chi_{c1}$. 

\section{Other exclusive decay modes}

\subsection{$B^0 \to J/\psi \pi^+ \pi^-$}

\begin{figure}[t]
\begin{center}
\epsfig{file=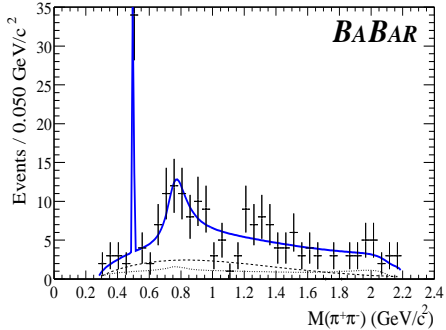,height=1.8in,width=2.4in}
 \caption{Distribution of the invariant mass M($\pi^+\pi^-$) for the
 $B$ candidates. The solid line is the result of the unbinned likelihood 
fit. The dotted (dashed) line represents the background from non-$J/\psi$
(inclusive-$J/\psi$) events.~\cite{Aubert:2002yk}.} 
\label{fig:jpipi}
\end{center}
\end{figure}

The BaBar collaboration has measured the branching fraction for the $B \to
J/\psi \pi^+ \pi^-$ decay~\cite{Aubert:2002yk}. The data set contains
approximately 56 million $B\bar{B}$ pairs produced at the $\Upsilon(4S)$ resonance with
the BaBar detector at the PEP-II asymmetric $e^+e^-$ collider. The $B \to
J/\psi \pi^+ \pi^-$ decay mode includes $J/\psi \rho^0$ and the
non-resonant $J/\psi \pi^+ \pi^-$ components. The invariant mass of the two pions,
$M(\pi^+ \pi^-)$, is plotted in Fig.~\ref{fig:jpipi}. The signal yield is
obtained by an unbinned maximum likelihood fit performed on the
invariant mass distribution. The branching fraction for $B^0 \to
J/\psi \pi^+ \pi^-$ is measured to be $(5.0\pm0.7\pm0.6)\times
10^{-5}$. It is about 20 times smaller than the
branching fraction for $B^0 \to J/\psi K^0$ because of the Cabibbo
suppression of $B \to J/\psi \pi^+ \pi^-$.

\subsection{$B \to J/\psi K^*$}

The $J/\psi K^*$ system has three helicity states and hence is a
mixture of the CP-even and CP-odd eigenstates. The full angular
analysis can determine the CP mix, which must be known to
measure $\rm{sin}2\phi_1$ when the decay $B^0 \to J/\psi K^{*0}$,
$K^{*0} \to K_S \pi^0$ is used. The angular analysis also provides a
test of the validity of the factorization hypothesis for B meson
decays to charmonium. 

\begin{figure}[ht]
\begin{center}
\epsfig{file=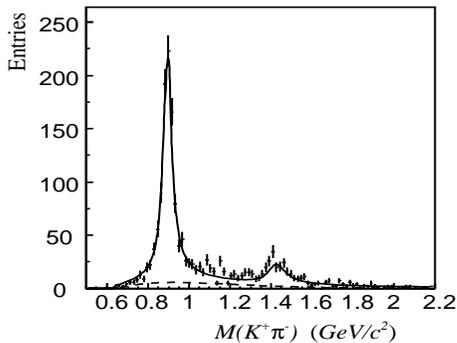,height=1.8in,width=2.4in}
 \caption{The $K^+\pi^-$ invariant mass distribution from Belle. The solid line
 is a fit to two Breit-Wigner functions corresponding to $K^*(892)$
 and $K^{*}_2(1430)$ with a background function (dashed line).} 
\label{fig:Mkpi}
\end{center}
\end{figure}

The branching fractions for $B^+ \to J/\psi K^{*+}$ and $B^0 \to
J/\psi K^{*0}$, where $K^{*+} \to K^+\pi^0$, $K_S\pi^+$ and $K^{*0}
\to K^+\pi^-$,$K_S\pi^0$, are listed in Table~\ref{tab:psikstar} for
comparison~\cite{Jessop:1997jk}~\cite{Aubert:2001pe}. The results are
all in good agreement. Figure~\ref{fig:Mkpi} also shows evidence for the
decay $B^0 \to J/\psi K^{*0}_2(1430)$. Some excess is observed in the
region between 1.1 $\rm{GeV}/c^2$ to 1.3 $\rm{GeV}/c^2$ in these
measurements. Its source is not fully understood.

\begin{figure}[ht]
\begin{center}
\epsfig{file=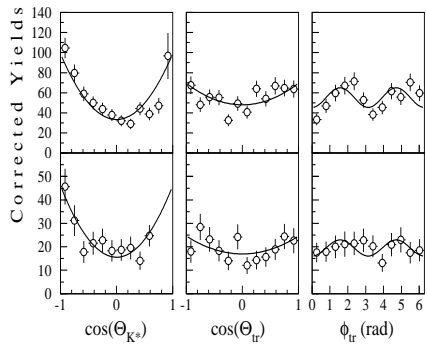,height=1.8in,width=2.2in}
 \caption{The background-subtracted angular distributions for the
 channels without (top) and without (bottom) a $\pi^0$. The curves
 correspond to the fit~\cite{Aubert:2001pe}.} 
\label{fig:angular}
\end{center}
\end{figure}

The decay amplitudes of $B \to J/\psi K^*$ are measured in the
transversity frame~\cite{Dunietz:1990cj} by fitting the angular distributions
in Fig.~\ref{fig:angular}. The results from various experiments are
compared in
Table~\ref{tab:amplitudes}~\cite{Jessop:1997jk}~\cite{Affolder:2000ec}~\cite{Aubert:2001pe}.
They are consistent. The value of
$|A_\perp|^2$, which corresponds to the CP-odd eigenstate, shows that
CP-even component dominates in the $B^0 \to J/\psi K^{*0}$, $K^{*0}
\to K_S \pi^0$ decay. The parameter $arg(A_\parallel)$ should be 0 or
$\pi$ in the factorization limit. It is shifted from $\pi$ in all four
measurements. However, the shift is not yet statistically significant
enough to draw a conclusion. 

\begin{table}[ht]
\begin{center}
\begin{tabular}{lll}  
\hline \hline 
Experiment & ${\cal{B}} (B^0 \to J/\psi K^{*0})$ ($\times 10^{-3}$) & ${\cal{B}} (B^+ \to
J/\psi K^{*+})$ ($\times 10^{-3}$) \\ 
 \hline 
CLEO   & $1.32\pm0.15\pm0.17$ & $1.41\pm0.20\pm0.24$  \\
BaBar  & $1.24\pm0.05\pm0.09$ & $1.37\pm0.09\pm0.11$ \\ 
Belle  & $1.29\pm0.05\pm0.13$ & $1.28\pm0.07\pm0.14$  \\ \hline\hline

\end{tabular}
\caption{The measured branching fractions for $B^+ \to J/\psi K^{*+}$
and $B^0 \to J/\psi K^{*0}$. }
\label{tab:psikstar}
\end{center}
\end{table}

\begin{table}[ht]
\begin{center}
\begin{tabular}{lcccc}  
\hline \hline 
Exp. & $|A_0|^2$ &  $|A_\perp|^2$& $arg(A_\parallel)$ (rad) & $arg(A_\perp)$ (rad) \\ 
 \hline 
CLEO   & $0.52\pm0.07\pm0.04$ & $0.16\pm0.08\pm0.04$ &
$3.00\pm0.37\pm0.04$ & $-0.11\pm0.46\pm0.03$ \\
CDF    & $0.59\pm0.06\pm0.01$ & $0.13^{+0.12}_{-0.09}\pm0.06$ & $2.2\pm0.5\pm0.1$ &$-0.6\pm0.5\pm0.1$ \\ 
BaBar  & $0.60\pm0.03\pm0.02$ & $0.16\pm0.03\pm0.01$ & $2.50\pm0.20\pm0.08$ & $-0.17\pm0.16\pm0.07$\\ 
Belle  & $0.62\pm0.02\pm0.03$ & $0.19\pm0.02\pm0.03$ & $2.83\pm0.19\pm0.08$ & $-0.09\pm0.13\pm0.06$\\ \hline\hline
\end{tabular}
\caption{The decay amplitudes in $B \to J/\psi K^*$. The first
errors are statistical and the second systematic.}
\label{tab:amplitudes}
\end{center}
\end{table}

\subsection{$B \to \eta_c K^{(*)}$}

The decay $B \to \eta_c K$ has the same quark level diagram as $B
\to J/\psi K$. However, unlike $J/\psi$, $\eta_c$ decays hadronically
rather than leptonically with rates of a few percent or less
for each channel. The decay modes $B^0 \to \eta_c K^0$, $\eta_c \to
K_S^0 K^- \pi^+$ and $K^+K^-\pi^0$ have been used to measure
$\rm{sin}2\phi_1$. Other decay channels of the $\eta_c$ may also be useful for
future CP measurements. 

BaBar has measured the branching fractions of $B^+ \to \eta_c K^+$ and
$B^0 \to \eta_c K^0$~\cite{Aubert:2002ht} using a  data sample containing 22.7 million
$B\bar{B}$ pairs. The $\eta_c$ is reconstructed in the decay
modes: $K_S^0 K^- \pi^+$, $K^+K^-\pi^0$, and $2(K^+K^-)$. They
observed statistically significant B meson signals in the $K_S^0 K^-
\pi^+$ and $K^+K^-\pi^0$ channels. They also observed exclusive
$\eta_c$ signals. 

\begin{figure}[t]
\begin{center}
\epsfig{file=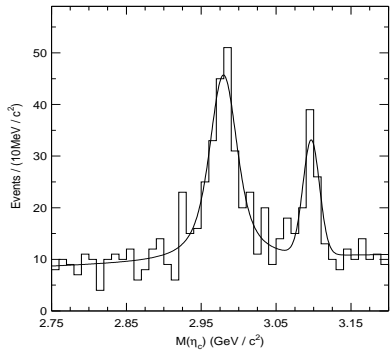,height=1.8in,width=2in}
 \caption{Candidate $M(\eta_c)$ invariant mass distribution for events
 in the $M_{bc}$ and $\Delta E$ signal region. Signals at the $\eta_c$
 and $J/\psi$ from $B \to \eta_c K$ and $B \to J/\psi K$ decays are visible.} 
\label{fig:metac2}
\end{center}
\end{figure}

Belle has measured the branching fraction of $B^+ \to \eta_c K^+$ and $B^0
\to \eta_c K^0$ using a  data sample containing 31.3 million
$B\bar{B}$ pairs. The $\eta_c$ is reconstructed in the decay
modes: $K_S^0 K^- \pi^+$, $K^+K^-\pi^0$, $K^{*0}K^-\pi^+$ and $p
\bar{p}$. We observed statistically significant B meson signals in
$K_S^0 K^- \pi^+$, $K^+K^-\pi^0$, $K^{*0}K^-\pi^+$ and in the $p
\bar{p}$ channel. Figure~\ref{fig:metac2} shows the
invariant mass of $\eta_c$ for events in the $M_{bc}$ and $\Delta E$
signal region. Fitting to a Breit-Wigner convolved with
the resolution determined from MC, we find a intrinsic width
$\Gamma(\eta_c)$ $=$ $29\pm8$ $\rm{MeV}$ and a mass of $M(\eta_c)$ $=$
$2979.6\pm2.3$ $\rm{MeV}$. The errors are statistical only. The
results are consistent with world averages~\cite{Groom:in} and the
CLEO result~\cite{Brandenburg:2000ry}. 

The $B$ branching fractions are quoted for the $\eta_c \to K_S^0 K^-
\pi^+$ and $\eta_c \to K^+K^-\pi^0$ modes only. The $\eta_c \to K_S^0 K^-
\pi^+$ mode is the most precisely and reliably measured mode, while the
branching fraction for the $\eta_c \to K^+K^-\pi^0$ mode is related by
isospin. The results are consistent with the CLEO
results~\cite{Edwards:2000bb} but more precise as shown in Table~\ref{tab:etac}.

\begin{table}[ht]
\begin{center}
\begin{tabular}{lll}  
\hline \hline 
Experiment & ${\cal{B}} (B^0 \to \eta_c K^{0})$ ($\times 10^{-3}$) & ${\cal{B}} (B^+ \to
\eta_c K^{+})$ ($\times 10^{-3}$) \\ 
 \hline 
CLEO   & $1.09^{+0.55}_{-0.42}\pm0.12\pm0.31$ & $0.69^{+0.26}_{-0.21}\pm0.08\pm0.20$ \\
BaBar  & $1.06\pm0.28\pm0.11\pm0.33$ & $1.50\pm0.19\pm0.15\pm0.46$ \\ 
Belle  & $1.23\pm0.23^{+0.12}_{-0.16}\pm0.38$ & $1.25\pm0.14^{+0.10}_{-0.12}\pm0.38$ \\ 
\hline\hline

\end{tabular}
\caption{The measured branching fractions for $B^+ \to \eta_c K^{+}$
and $B^0 \to \eta_c K^{0}$. The last errors come from the uncertainty
in the $\eta_c$ branching fraction}
\label{tab:etac}
\end{center}
\end{table}

Belle has observed the decay mode $B^0 \to \eta_c K^{*0}$ for the first
time. The $K^{*0}$ is reconstructed in the $K^-\pi^+$ channel and the
$\eta_c$ in the $K_S^0 K^- \pi^+$ mode. To remove the $B \bar{B}$
background, we apply vetoes to events consistent with $J/\psi \to K_S^0
K \pi$, $\chi_{c1} \to K_S^0 K \pi$ and $D_s \to K^+K^-\pi$. A fit to
the $M_{bc}$ spectrum yields a signal of $33.7\pm6.7$ events with a statistical
significance of $7.7\sigma$. The branching fraction for $B^0 \to
\eta_c K^{*0}$ is found to be
($1.62\pm0.32^{+0.24}_{-0.34}\pm0.50)\times 10^{-3}$. The ratio
$R_{\eta_c}$ = ${\cal{B}}(B^0 \to \eta_c K^{*0})/{\cal{B}}(B^0 \to \eta_c K^{0})$
is measured to be $1.33\pm0.36^{+0.29}_{-0.40}$. This result is
somewhat higher than the theoretical prediction of Gourdin, Keum and
Pham of 0.78~\cite{Gourdin:1994xx}.

\begin{figure}[t]
\begin{center}
\epsfig{file=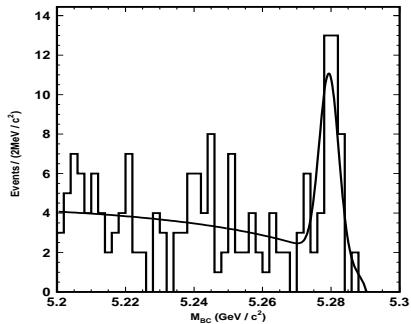,height=1.8in,width=2.2in}
\caption{ The $M_{bc}$ distribution of $B^0 \to \eta_c K^{*0}$ candidates.}  
\label{fig:mbc}
\end{center}
\end{figure}

\subsection{Exclusive $B \to J/\psi$, $\psi(2S)$ and $\chi_{c1}$}

Table~\ref{tab:others} lists branching fractions for two-body $B$
meson decays to $J/\psi$, $\psi(2S)$ and $\chi_{c1}$ with a kaon or
pion~\cite{Groom:in}~\cite{Richichi:2000ca}~\cite{Abe:2001na}~\cite{Abe:2001wa}~\cite{Aubert:2001xs}~\cite{Aubert:2002yv}.
They are in good agreement with previous measurements but more
precise. The decay modes $ B^0 \to \chi_{c1}
K^{*0}$,  $B^+ \to J/\psi K_1^+(1270)$ and $B^0
\to J/\psi K_1^0(1270)$ have been observed for the first time. 

\begin{table}[ht]
\begin{center}
\begin{tabular}{llll}  
\hline\hline
Decay mode & Previous & BaBar &  Belle \\ 
& ($\times 10^{-4}$) & ($\times 10^{-4}$) & ($\times 10^{-4}$) \\
 \hline
$\modeI$   &$10.0\pm1.0$ & $10.1\pm0.3\pm0.5$ & $10.1\pm0.3\pm0.8$  \\
$\modeII$  &$9.6\pm0.9$ & $8.3\pm0.4\pm0.5$ & $7.7\pm0.4\pm0.7$ \\ 
$\modeIII$ &       &         & $18.0\pm3.4\pm3.9$  \\
$\modeIV$  &      &         &  $13.0\pm3.4\pm3.1$  \\ \hline
$\modeV$   &$5.8\pm1.0$ & $6.4\pm0.5\pm0.8$ & $6.7\pm0.6\pm0.7$ (a) \\
           &      &         &  $5.7\pm0.5\pm0.8$ (b)   \\ 
$\modeVI$  &$5.0\pm1.3$ & $6.9\pm1.1\pm1.1$ & $6.0\pm1.1\pm0.7$ (a)    \\
           &      &         &  $7.2\pm1.1\pm1.1$  (b)   \\ \hline
$\modeVII$  &$10.0\pm4.0$  & $7.5\pm0.8\pm0.8$ & $6.1\pm0.6\pm0.6$   \\ 
$\modeVIII$ &$3.9^{+1.9}_{-1.4}$ & $5.4\pm1.4\pm1.1$ & $3.1\pm0.9\pm0.4$ \\
$\modeIX$   &      &  $4.8\pm1.4\pm0.9$ &   \\ \hline
$\modeX$    &$0.51\pm0.15$ & $0.39\pm0.09$ & $0.52\pm0.07\pm0.07$   \\ 
$\modeXI$   &$0.25^{+0.11}_{-0.09}$& $0.20\pm0.06\pm0.02$ &
$0.24\pm0.06\pm0.02$ \\ \hline\hline 
\end{tabular}
\caption{Measured branching fractions. (a) $\modeXII$ (b) $\modeXIII$. }
\label{tab:others}
\end{center}
\end{table}

\section{Observation of $\eta_c(2S)$ meson}

The $\eta_c(2S)$ meson has not been experimentally well
established. The Crystal Ball group reported possible evidence for
the $\eta_c(2S)$ meson with a mass of $3594\pm5$
$\rm{MeV}$~\cite{Edwards:1981mq}. The result has
not been confirmed by the subsequent experiments. 

\begin{figure}[t]
\begin{center}
\epsfig{file=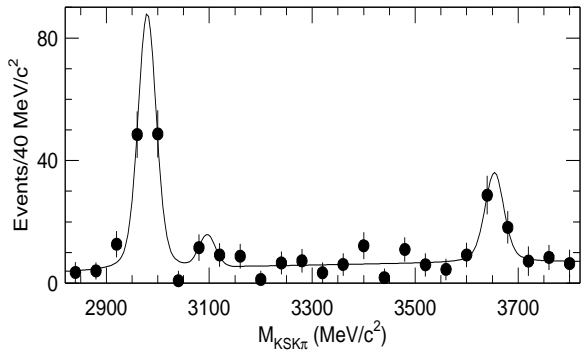,height=1.8in,width=3in}
 \caption{ The signal yields for each $K_SK\pi$ bin. The curve is the
 result of the fit.}
\label{fig:slice}
\end{center}
\end{figure}

Using a data set that contains 44.1 million $B\bar{B}$ pairs, Belle has
searched for the $\eta_c(2S)$ meson produced via the exclusive decays
$B^+ \to \eta_c(2S) K^+$ and $B^0 \to \eta_c(2S) K^0$ where
$\eta_c(2S) \to K_S K^-\pi^+$. To remove backgrounds from $B \to D(D_s)X$ and
$B\to K^*(890)K$ decays, $D$, $D_s$ and $K^*$ vetoes are
applied. The $M_{bc}$ and $\Delta E$ distributions are
plotted for twenty-five $M_{K_SK\pi}$ bins. Clear $B$ meson signals are
seen in the bins corresponding to the $\eta_c$ and near the expected mass of
the $\eta_c(2S)$. The signal yields extracted from the simultaneous fits to the
$M_{bc}$ and $\Delta E$ distributions are plotted in
Fig.~\ref{fig:slice}. A clear peak is seen around $3.65$
$\rm{GeV}/c^2$ and identified as $\eta_c(2S)$. The distribution is fit
to two Breit-Wigner functions for the $\eta_c$ and $\eta_c(2S)$ respectively, a
Gaussian for the $J/\psi$, and a second-order polynomial for the
non-resonant contribution. These functions are convolved with a
Gaussian resolution function determined from MC. The fit value for the
mass is $3654\pm6\pm8$ $\rm{MeV}/c^2$. The $90\%$ confidence level
upper limit for the intrinc width is 55 $\rm{MeV}/c^2$. The
results are consistent with expectations of heavy-quark
potential models. The ratio of product branching fractions
for the $\eta_c$ and $\eta_c(2S)$ is also measured to be 
\[ \frac{{\cal{B}}(B \to \eta_c(2S)K){\cal{B}}(\eta_c(2S) \to K_SK^-\pi^+)}{{\cal{B}}(B \to
\eta_c K){\cal{B}}(\eta_c \to K_SK^-\pi^+)} = 0.38\pm0.12\pm0.05. \]
%

\section{Summary}

With the large $B \bar{B}$ data sets accumulated at the B-factories, we
have improved the measurements of the branching fractions for the
decays $B \to J/\psi K^{(*)}$, $B \to \psi(2S) K$, $B \to \chi_{c1} K$
and $B \to J/\psi \pi$. The decays modes $B^0 \to \chi_{c1} K^{*0}$, $B^+
\to \chi_{c0} K^+$, $B \to \chi_{c2} X$, $B^0 \to \eta_{c} K^{*0}$ and
$B \to J/\psi K_1(1270)$ have been observed for the first time. The
branching fractions for the non-factorizable decays $B^+ \to \chi_{c0} K^+$
and $B \to \chi_{c2} X$ are comparable to the factorizable decays
$B^+ \to J/\psi K^+$ and $B \to \chi_{c1} X$. Belle has observed the
$\eta_c(2S)$ meson. Its properties are consistent with
expectations of the heavy-quark potential models.

\end{document}